\documentclass[twocolumn]{aastex63}

\usepackage{graphicx}
\usepackage{amsmath}
\usepackage{mathtools}
\usepackage{amssymb}

\newcommand{\code}[1]{\texttt{#1}}
\newcommand{\urgent}[1]{{\textcolor{black}{#1}}}
\newcommand{\moreurgent}[1]{{\textcolor{black}{#1}}}
\newcommand{\veryurgent}[1]{{\textcolor{black}{#1}}}

\usepackage{graphicx}

\accepted{February 21, 2023}

\shorttitle{The Stellar Halo of M94}
\shortauthors{Gozman et al.}

\graphicspath{{./}{figures/}}

\begin{document}

\title{Saying Hallo to M94's Stellar Halo: Investigating the Accretion History of the Largest Pseudobulge Host in the Local Universe}

\correspondingauthor{Katya Gozman}
\email{kgozman@umich.edu}

\author[0000-0003-2294-4187]{Katya Gozman}
\affiliation{Department of Astronomy, University of Michigan, 1085 S. University Ave, Ann Arbor, MI, 48109-1107, USA}

\author[0000-0002-5564-9873]{Eric F.\ Bell}
\affiliation{Department of Astronomy, University of Michigan, 1085 S. University Ave, Ann Arbor, MI, 48109-1107, USA}

\author[0000-0003-2599-7524]{Adam Smercina}
\affiliation{Astronomy Department, University of Washington, Box 351580, U.W. Seattle, WA, 98195-1580, USA}

\author{Paul Price}
\affiliation{Department of Astrophysical Sciences, Princeton University, Princeton, NJ, 08544, USA}

\author[0000-0001-6380-010X]{Jeremy Bailin}
\affiliation{Department of Physics and Astronomy, University of Alabama, Box 870324, Tuscaloosa, AL, 35487-0324, USA}

\author[0000-0001-6982-4081]{Roelof \urgent{S.} de Jong}
\affiliation{Leibniz-Institut f\"{u}r Astrophysik Potsdam (AIP), An der Sternwarte 16, D-14482 Potsdam, Germany}

\author[0000-0001-9269-8167]{Richard D'Souza}
\affiliation{Vatican Observatory, Specola Vaticana, I-00120, Vatican City State, Italy}

\author[0000-0002-2502-0070]{In Sung Jang}
\affiliation{Department of Astronomy and Astrophysics, University of Chicago, Chicago, IL, 60637, USA}

\author[0000-0003-2325-9616]{Antonela Monachesi}
\affiliation{Instituto de Investigaci\'{o}n Multidisciplinar en Ciencia y Tecnolog\'{i}a, Universidad de La Serena, R\'{a}ul Bitr\'{a}n 1305, La Serena, Chile}
\affiliation{Departamento de Astronom\'{i}a, Universidad de La Serena, Av. Juan Cisternas 1200 N, La Serena, Chile}

\author[0000-0002-0558-0521]{Colin Slater}
\affiliation{Astronomy Department, University of Washington, Box 351580, U.W. Seattle, WA 98195-1580}

\begin{abstract}
It is not yet settled how the combination of secular processes and merging gives rise to the bulges and pseudobulges of galaxies. The nearby ($D\sim$4.2\,Mpc) disk galaxy M94 (NGC 4736) has the largest pseudobulge in the local universe, and offers a unique opportunity for investigating the role of merging in the formation of its pseudobulge. We present a first ever look at the M94's stellar halo, which we expect to contain a fossil record of M94’s past mergers. Using Subaru’s Hyper Suprime-Cam, we resolve and identify red giant branch (RGB) stars in M94’s halo, finding two distinct populations. After correcting for completeness through artificial star tests, we can measure the radial profile of each RGB population. The metal-rich RGB stars show an unbroken exponential profile to a radius of 30\,kpc that is a clear continuation of M94's outer disk.  M94's metal poor stellar halo is detectable over a wider area and clearly separates from its metal-rich disk. By integrating the halo density profile, we infer a total accreted \urgent{stellar} mass of  $\sim 2.8 \times 10^8 M_\odot$, with a median metallicity of [M/H]$=-$1.4. \veryurgent{This indicates that M94’s most-massive past merger was with a galaxy similar to, or less massive than, the Small Magellanic Cloud. Few nearby galaxies have had such a low-mass dominant merger; therefore we suggest that M94's pseudobulge was not significantly impacted by merging.}

\end{abstract}


\section{Introduction}\label{sec:intro}

Galaxy formation is driven by both secular and hierarchical processes: the former being the formation of stars and structures that happens internally in a galaxy as it evolves over time (in situ processes), and the latter being accreted through mergers and interactions with other galaxies (ex situ processes; e.g., \citet[][]{kormendy1979, barneshernquist1992, conselice2008structures}). Though past studies \citep[e.g.,][]{KK2004} have postulated that the nonsecular processes of hierarchical clustering and violent mergers were dominant in the early universe, giving way to slow and secular processes at later times, we now know that the evolution of galaxy features seen today arises from a combination of and interplay between these two processes, both of which are important in shaping the diverse morphologies of today's galaxies \citep[e.g.,][]{hopkins2010, somerville2015physical}.

Secular evolution is the slow and steady evolution of a galaxy through either long-term interactions with its environment or internal process with timescales that are long compared to the dynamical timescale of the galaxy. Galaxies accumulate gas from external accretion and infall, which is then used as a catalyst for secular processes and the formation of disk instabilities and tidal interactions (\citealt{combes2009secular}). On the other hand, hierarchical evolution happens on dynamical timescales through events like mergers and satellite accretion, wherein galaxies are built out of mergers with smaller galaxies \citep{whiterees1978, barneshernquist1992, bullockjohnson, purcell2007, cooper2010}. Mergers should therefore be important drivers of galaxy evolution. They can help shape a galaxy's formation, morphological features such as bulges and disks, metallicity and color gradients, kinematics, and stellar populations \citep[e.g.,][]{kaufmann1999, deanson2015, brookeschristiansen2016, dsouzabell2018b, gallart2019, gargulio2019}.

Both of these impetuses have been predicted to result in different morphological evolutions for a galaxy. Disk instabilities from secular evolution can trigger gas to cascade into the center of the galaxy and impart angular momentum into the system. This high density of gas can fuel star formation and nuclear starbursts (\citealt{kormendy2007}), as well as form other structures such as spiral disks and bars. Bars also commonly drive gas inward and grow by transferring angular momentum to the outer disk. Even if a galaxy has a weak bar (or has an oval disk), a nonaxisymmetric potential is still created and the evolution of the central part of the galaxy is similar to barred galaxies (\citealt{KK2004}). 

The infalling gas can then create a structure known as a pseudobulge. Pseudobulges are the dense, flat, inner region of a galaxy that have retained some of their disk-like properties (\citealt{athanassoula2005}). They have more rotation than random-motion-dominated bulges, have active ongoing star formation with younger stellar populations that can fuel their growth, smaller S\'{e}rsic indices, and they do not follow the Faber-Jackson relation or fall on the fundamental plane (\citealt{KK2004}, \citealt{fisherdrory}, \citealt{KormendyDrory_bulgelessgiants_2010}). 

Alternatively, it is widely believed that hierarchical evolution is a catalyst for the formation of classical bulges (\citealt{hopkins2010}, \citealt{brookeschristiansen2016}), the bright center structure in some galaxies, historically called ``ellipticals sitting in the middle of disks" (\citealt{renzini1999fgb..conf....9R}). But this is not as clear cut as it seems: the formation of bulges during mergers is nearly inevitable in galaxy formation simulations \citep{KormendyDrory_bulgelessgiants_2010}, leading to an overrepresentation of bulge-dominated galaxies, relative to their observed prevalence in the nearby universe. The evolution of galactic features is an incredibly complex and dynamic process and one that can be influenced by both secular and nonsecular evolution at various stages \citep{brookeschristiansen2016, gargulio2019, Izquierdo_Villalba_2019}.

In order to learn about and constrain the creation physics of galactic bulge structures, we turn to the disk galaxy M94 (NGC 4736, D = 4.2 Mpc, GHOSTS (Galaxy Halos, Outer disks, Substructure, Thick disks, and Star clusters), \citealt{ghosts}). \urgent{M94 has a stellar mass of $\sim5.37 \times 10^{10} M_\odot$ \citep{Karachentsev_dist}, thereby making it a Milky Way-stellar mass galaxy assuming the Milky Way has a stellar mass of $6.08 \pm 1.14 \times 10^{10} M\odot$ \citep[][]{Licquia2015}. At its center, M94 has an oval distortion and weak bar in its center that contains low-ionization nuclear emission-line region (LINER) emission \citep{Trujillo_2009, Watkins2016}. Beyond, it hosts a exponentially declining inner disk that flattens out, hence classified as an antitruncated disk \citep{Trujillo_2009}, which \cite{younger2007} states can arise from a minor merger in certain cases. It has an intensely star-forming inner ring at around 50'' ($\sim 1 \text{ kpc}$;
\citet{Watkins2016}). Outside this, \cite{Trujillo_2009} did an intensive, multiwavelength study of M94's outer disk and found that in the UV and infrared, the outer parts of the galaxy shows a spiral arm structure with increased star formation as compared to the inner disk. While they favor a secular evolution explanation for this bright star-forming outer disk, they say they cannot rule out a scenario in which this extended disk originates from satellite accretion. In the optical, this spiral arm structure appears as an outer star-forming ring at 4 kpc and has some asymmetric features inside of 10 kpc \citep[][]{Watkins2016}. Studies such as by \cite{herrmannetal2009_PNe} have used planetary nebula kinematics to show that M94's disk is flared at its outer parts where older stellar populations reside, which may be due to past interactions.} 

\urgent{While estimates of M94's rotational velocity are $\sim$120 km/s \citep{deBlok2008}, considerably lower than the Milky Way's and considerably lower than would be expected for a galaxy of M94's stellar mass, the combination of M94's fairly face-on orientation and warped disk make its rotation velocity \moreurgent{and inclination} highly uncertain. Indeed, multiple studies have shown that M94 has large, noncircular motions throughout its disk \citep[][]{mollenhoff1995, wong2000noncircgas,deBlok2008, trachternach2008, walter2008_thingsnoncirc, Watkins2016} and therefore one should use caution when interpreting its rotation curve or using its measured circular velocities for comparison with other galaxies. \cite{deBlok2008} specifically warn against using M94's velocity profile to infer its total mass given the dominant impact of these noncircular motions. Accordingly, when we compare M94 with other galaxies, we compare it with galaxies with comparable stellar masses, irrespective of their published rotational velocities.} \moreurgent{Curiously enough, the presence of a warp in M94's disk and its large noncircular motions are features that have been attributed to interactions in other galaxies such as the Milky Way \citep{lopezcorredoira2002, bailin2003, lopezcorredoira2006, poggio_MW_warp2020}. If M94 does not have evidence of a large star-rich merger, this would indicate that such morphological features can arise from nonmerger impetuses such as secular evolution or gas accretion from the surrounding medium.}

M94 is unique in that it has the largest pseudobulge in our local universe, containing about 50\% of its stellar mass (\citealt{Trujillo_2009}). This pseudobulge is so concentrated and massive that this could only be caused by a significant loss of angular momentum by the system. Though M94 has been cited as a prime example of a pseudobulge galaxy driven by secular evolution (\citealt{KK2004}), M94's pseudobulge is so large, and bulge formation processes are sufficiently complex (\citealt{stinson, brookeschristiansen2016}) that mergers could have contributed to its formation. Simulations have shown that there are examples of pseudobulges forming and surviving through active processes such as mergers (\citealt{combes2005}, \citealt{kormndyfisher2008}, \urgent{\citealt{governato2009}, \citealt{moster2010}}, \citealt{keselmannusser}). \urgent{In fact, \cite{barway2020} discovered a pseudobulge in the collisional ring galaxy Cartwheel, showing that pseudobulges can survive drop-through collisions in certain situations. Studies by \cite{elichemoral2006} and \cite{elichemoral2011} found that pseudobulge signatures can be created by satellite accretion, where the infalling satellite creates a disk instability that drives gas inward.} \cite{guedes2013} \urgent{also} showed that a pseudobulge could form quickly at high redshift from tidal interactions or mergers on dynamical, not secular, timescales, and other studies such as by \cite{Okamoto_2012} have found that under certain simulated conditions the main formation mechanism for a pseudobulge was starburst activity at early times (completed at $z < 2$). 

With this in mind, we use M94 as a testbed for galaxy evolution and aim to:

\begin{enumerate}
    \item Characterize M94's stellar halo; and
    \item Answer the question: could M94's dominant merger have played a role in the creation of its large pseudobulge, or was it formed due to other processes such as secular evolution?
\end{enumerate}

Studying M94's merger history and characterizing its stellar halo can shed light on this question and also give us insight into how a disk galaxy's major merger can affect its internal structure.

\begin{figure*}[t]
    \centering
    \includegraphics[width=\linewidth]{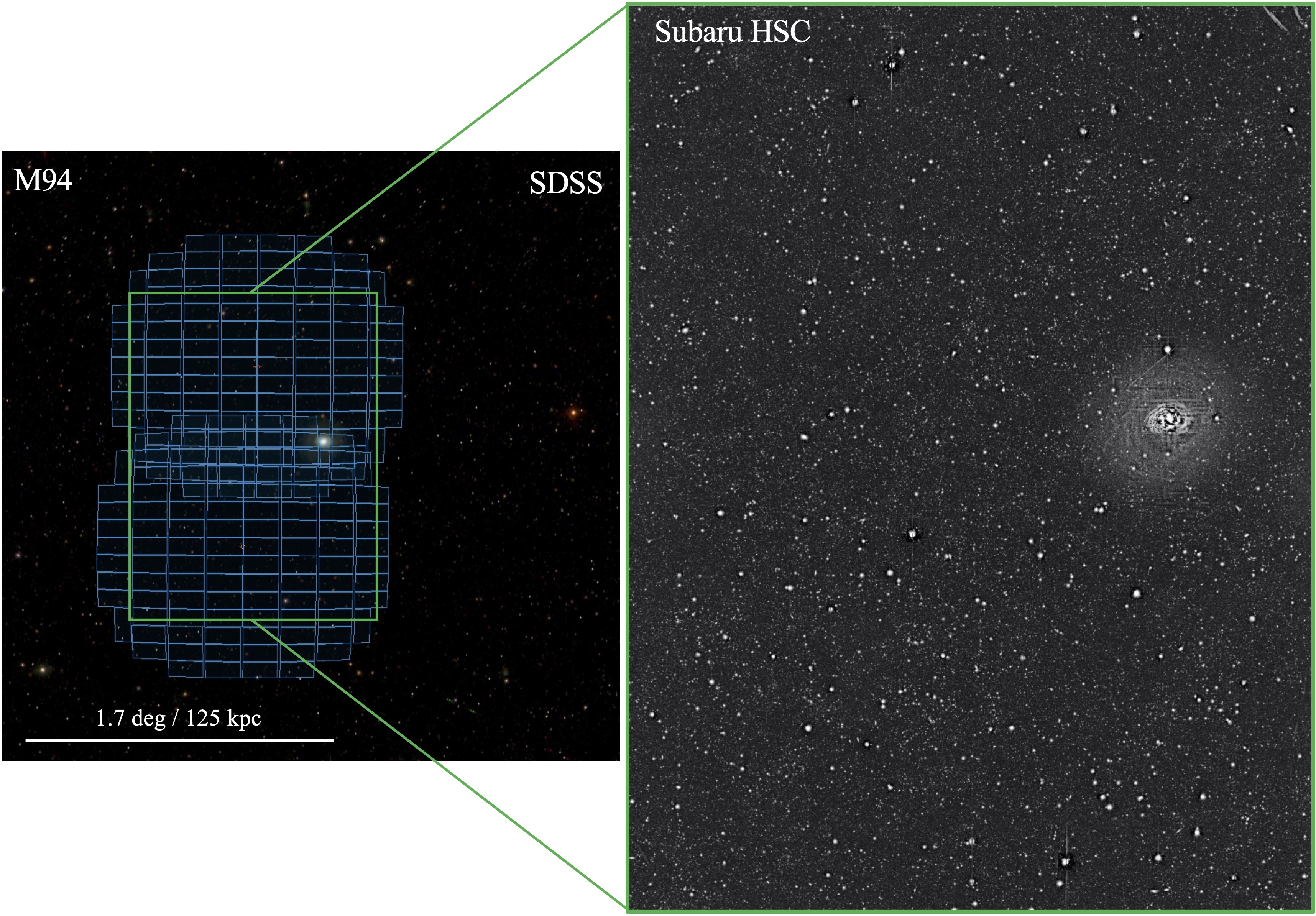}
    \caption{\textit{Left}: a 3\fdg5\,$\times$\,3\fdg5 view of M94 and its surroundings, taken from SDSS, with the field of view of our two deep Subaru HSC pointings overlaid (using the \textit{Aladin Sky Atlas}). \textit{Right}: an $r$-band mosaic of the central 1\fdg33\,$\times$\,1\fdg76 of observations taken in our two Subaru HSC pointings. The image has been logarithmically scaled. M94's bright outer ring feature is visible as diffuse brightness, and we resolve the outermost parts of this ring, as well as M94's diffuse halo, into stars.}
    \label{fig:m94-mosaic}
\end{figure*}

That said, detecting and learning about the past mergers of galaxies is a difficult task. In the currently accepted $\Lambda$-cold dark matter ($\Lambda$CDM) cosmological framework, galaxies form through the process of hierarchical assembly wherein a large DM halo and the galaxy associated with it grows through many mergers with smaller objects over time (\citealt{whiterees1978, bullock2001}). Even though an ongoing active merger often produces very distinctive and observable tidal features, the observational markers of mergers blend in with the main galaxy on short timescales as the gas disperses throughout the main body and the stars undergo phase mixing \citep{shapiro1972, bullockjohnson}. This makes identifying the properties of a past merger a challenging task.

Fortunately, not all information is lost. During a merger, the \urgent{infalling} galaxy will deposit material into the main galaxy, and its disrupted stars will spread out and form a diffuse population of stars surrounding the disk of the main galaxy that we call the stellar halo (\citealt{bullockjohnson}), a remnant of its hierarchical and collisionless merger history. Therefore, information about past accretions is encoded into the stellar halos of galaxies, making them useful fossils of past mergers (\citealt{ bullockjohnson, purcell2007, cooper2010, Bell2017, DSouzaBell2018a, monachesi2019}). Theory and observations suggest that the halos of Milky Way-\urgent{stellar} mass galaxies predominantly manifest the properties of the most-massive galaxies that have previously merged with them (\citealt{deluciahelmi2008, deanson2015, DSouzaBell2018a, monachesi2019, Smercina2020}). \urgent{Stellar halos are have low surface brightnesses, so it is critical to resolve red giant branch (RGB) stars, which trace the underlying faint stellar population and are luminous enough to be detected even outside the Local Group \citep{ghosts}, unlike Main Sequence (MS) stars}. \urgent{Deriving }the mass, metallicity, and structure of a galaxy's halo \urgent{using RGB stars }enables us to infer its most dominant merger event. 

This insight has helped to investigate the impacts of mergers by using the stellar halos of the Milky Way (MW) and M31. Studies of the Milky Way's halo have led to the discovery that a large interaction with a $10^9 M_\odot$ galaxy around nine billion years ago (\citealt{ belokurov2018,helmi2018}) coincides with the formation of the Milky Way's thick disk (\citealt{gallart2019}), indicating an association between mergers of a characteristic size and thick disk formation. An investigation of the halo of our nearest neighbor, M31, showed that its most-massive merger thickened its disk (\citealt{dsouzabell2018b, hammer2018}) but failed to destroy it, causing a burst of star formation (\citealt{williams2017phat}), while bringing in many of M31's satellites (\citealt{weiscz2019, dsouzabell2021}). 

Cognizant that mergers may be important in shaping galaxy morphology \citep{somerville2015physical,brookeschristiansen2016} and in delivering satellite galaxies, \citep[][]{Deason2015, patel2017, dsouzabell2018b}, multiple approaches are used to measure and characterize the stellar halos that can constrain merger histories. The least expensive of these are deep imaging surveys (\citealt{martinezdelgado2010, vandokkum2014,Merritt2016, Watkins2016}). Resolved stellar population studies with Hubble Space Telescope (HST) pencil-beam observations (GHOSTS) or ground-based resolved-star wide-field imaging (\citealt{pandas,okamoto2015, cron2016, Smercina2020}) are more observationally expensive, but can reach fainter surface brightness limits, give stellar population information (metallicities, insight into star formation histories, etc.) and do not suffer from contamination from galactic cirrus or scattered light. 

In this paper, we present observations of the resolved stellar population of M94 using the Subaru's Hyper Suprime-Cam (HSC) in three passbands. Three passbands enable us to both remove contaminants and measure the halo to a greater depth and lower surface brightness. HSC affords us both the high sensitivity and large coverage area we need in order to be able to probe resolved stellar populations. We take an unprecedented wide-field, sensitive look at M94's stellar halo, and infer the \urgent{stellar} mass of its most dominant major merger in order to evaluate whether it could have played a role in the formation of M94's pseudobulge.

\section{Observations} \label{sec:obs}

Resolved stellar imaging of M94 was carried out during two nights (2017 20--21 April) using Subaru's HSC (\citealt{HSC}) through the Gemini--Subaru exchange program (PI: Smercina, NOAO 2017A-0312). Two fields of $\sim$1.5 square degrees each were surveyed in the $g$-, $r$-, and $i$ bands, with a depth of $\sim2$\,hr per band per pointing. The image scale of HSC is 1' per 1.22 kpc. Figure \ref{fig:m94-mosaic} shows an image of M94 from the Sloan Digital Sky Survey (SDSS), with the two HSC fields superimposed on top to show the full field of view. The combination of depth and image quality possible with HSC permit resolving the brightest tip of the red giant branch \urgent{(TRGB)} stars at the distance of M94, giving access to both stellar population information (e.g., metallicities from RGB colors; e.g., \citealt{Monachesi2016}) and extremely sensitive inferred surface brightness limits \citep{Smercina2020}. Information about the observation cadence, filters, integration time, and 50\% completeness limiting magnitudes is given in Table \ref{tab:obs}.

\begin{deluxetable}{cccc}

\tablecaption{ Observations \label{tab:obs}}

\tablehead{
\colhead{Band} & 
\colhead{Exp Time (s)} & 
\colhead{\# of Exposures} &
\colhead{50\% Completeness Mag}
}

\startdata 
$g$ & 7200 & 5 & 27.7 \\
$r$ & 6200 & 11 & 26.9 \\
$i$ & 9000 & 27 & 26.3 \\
\enddata

\tablecomments{Observation details for the Subaru HSC for each band in the program. The last column is the 50\% completeness limit for recovering sources in each band.}

\end{deluxetable}

The data were reduced and calibrated, and photometry was performed using the HSC optical imaging pipeline \citep{HSC_pipeline}. The pipeline performs photometric and astrometric calibration using the Pan-STARRS1 catalog \citep{PS1}, reporting the final magnitudes in the HSC natural system. For each source, we perform background subtraction by measuring the clipped mean of deblended pixels within an annulus that extends from 7 to 15 times the point-spread function (PSF) $\sigma$, and we subtract that times the effective area of the aperture from the source flux measurement.
We extract PSF photometry that has been corrected to match aperture photometry with a 12 pixel radius. Sources are detected in all three bands, although $i$ band is prioritized to determine reference positions for forced photometry. Forced photometry is then performed on sources in the $gri$\ coadded image stacks.

All magnitudes were corrected for (very modest) galactic extinction of an average E(B-V) = 0.01 following \citet{SFD98} and adopting the updated transformations from inferred reddening to extinction favored by \citet{Schlafly2011}. The depths of the images were nearly uniform across the two fields, yielding extinction-corrected $\sim 5\sigma$ point-source detection limits of $g\,{=}\,28.1$, $r\,{=}\,27.5$, and $i\,{=}\,27.0$. See \cite{HSC_pipeline} for an in-depth discussion of the photometric uncertainties output by the \texttt{HSC pipeline}. The seeing was relatively stable, resulting in consistent point-source sizes of $\sim$0\farcs75. 

\section{Data and Methods} \label{sec:datamethods}

\subsection{Stellar Selection}

Resolved stellar populations are powerful tools for measuring halo properties, but at the distance of M94, old low\urgent{-} and intermediate\urgent{-}mass MS stars, containing the bulk of the stellar mass, are too faint to detect at the depths of HSC. Therefore, we rely on measuring \urgent{RGB} stars whose metallicities are a strong function of their color \citep{streich}. They are also a more luminous population, hence we are able to resolve RGB stars within two magnitudes of the TRGB, which is 2-3 orders of magnitude brighter than their MS counterparts and still trace the underlying stellar population.

\begin{figure*}[!h]
    \centering
    \includegraphics[width=0.98\textwidth]{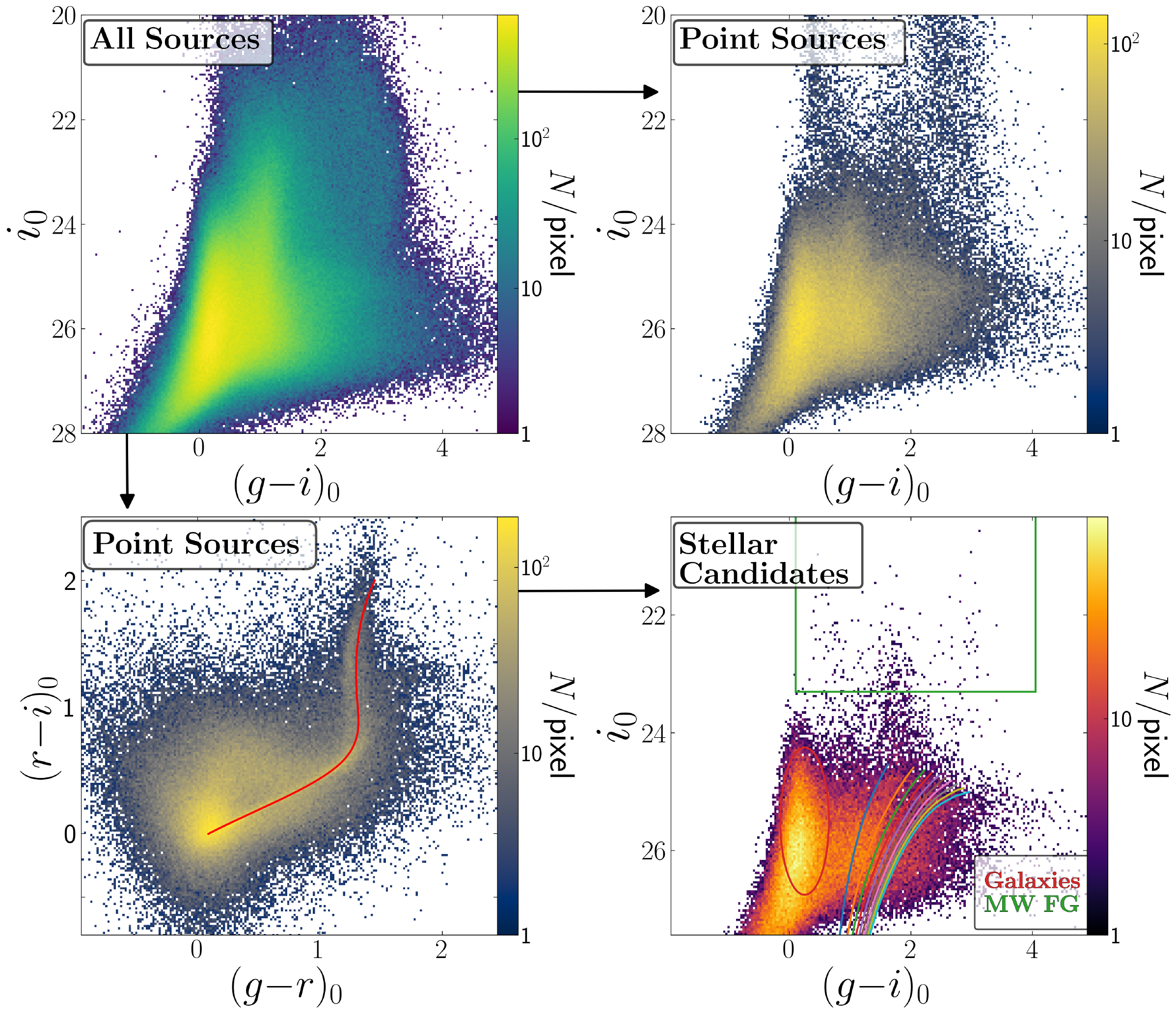}
    \caption{\textit{Top Left}: a Hess CMD of all sources in M94, as recovered by the HSC processing pipeline. \textit{Top Right}: a Hess CMD of only sources that met our criteria for being a point or compact source, as stated in Section \ref{sec:datamethods}. \textit{Bottom Left}: a color-color diagram of point sources. The red curve shows the defined stellar locus. \textit{Bottom Right}: a Hess CMD of sources that meet our selection criteria for being stellar candidates: being a point or compact source and lying within a certain quadrature-added error of the stellar locus. The green box shows the approximate area where the Milky Way foreground is located and the red ellipse is an area populated by unresolved contaminating background galaxies. Overlaid are also a set of ten isochrones (PARSEC; \citealt{bressan_parsec}) of age 10 Gyr with metallicities evenly spaced from $Z = 0.0002$ to 0.006. Figure template is taken from \cite{Smercina2020}.}
    \label{fig:fourpanel}
\end{figure*}

Unfortunately, at the depth achieved by HSC, resolved stellar sources are contaminated by galaxies because HSC can detect, but not resolve, distant galaxies (see Figure \ref{fig:fourpanel} top left), so it is important to have a systematic method to achieve star--galaxy separation. We first cut sources by FWHM with the criterion FWHM $<$\ 0\farcs75 to select only point and compact sources. This is shown at the top right of Figure \ref{fig:fourpanel}, which depicts a color-magnitude diagram \urgent{(CMD)} of all unresolved `point' sources in our sample. In shallower datasets, such cuts are sufficient owing both to the relatively lower number of galaxies and their typically larger FWHMs at brighter magnitudes. In our deep dataset, compact and unresolved galaxies {\it still} outnumber stars, being particularly apparent at colors $g{-}i \sim 0$.

\urgent{Stars occupy an empirical `stellar locus' in color-color diagrams that that can be distinct from galaxies.} We use the $g{-}r$/$r{-}i$\ stellar locus \citep[e.g.,][]{covey2007,ivezic2007,high2009,davenport2014} to select stars in our survey and remove contaminants like galaxies and quasars. This is shown as the red line in Figure \ref{fig:fourpanel} at the bottom left. We chose sources that are within their photometric $\sqrt{\sigma_{g-r}^2 + \sigma_{r-i}^2}$ errors of this line. This leaves us with our final list of stellar candidates, whose CMD is shown at the bottom right of Figure \ref{fig:fourpanel}. \urgent{The RGB sequence} is clearer in this CMD, and unresolved galaxies with $g-i \sim 1$ are much reduced in number (compare the top-right and bottom-right panels of Fig.\ \ref{fig:fourpanel}), but compact galaxies with $g-i \sim 0$ are still numerous, primarily because it is at this point where distant background galaxies and stars have similar $gri$ colors.

Using this final selection, we then narrow our focus to just the region around the RGB. 

\subsection{RGB Regions} \label{sec:rgbregions}

We wish to obtain a more robust color distribution that will better reflect the metallicity of the population, so we calculate the `Q-color' of candidate RGB stars following \citet{Monachesi2016}. The intention of this metric is to give the color of an RGB star, referenced to a magnitude 0.5 mag below the TRGB. We achieve this by rotating the (color, TRGB magnitude) point of (2,25.036) around a $-22^{\circ}$ angle, making the metal-poor RGB vertical. 

\begin{figure}[!h]
    \includegraphics[width=84 mm]{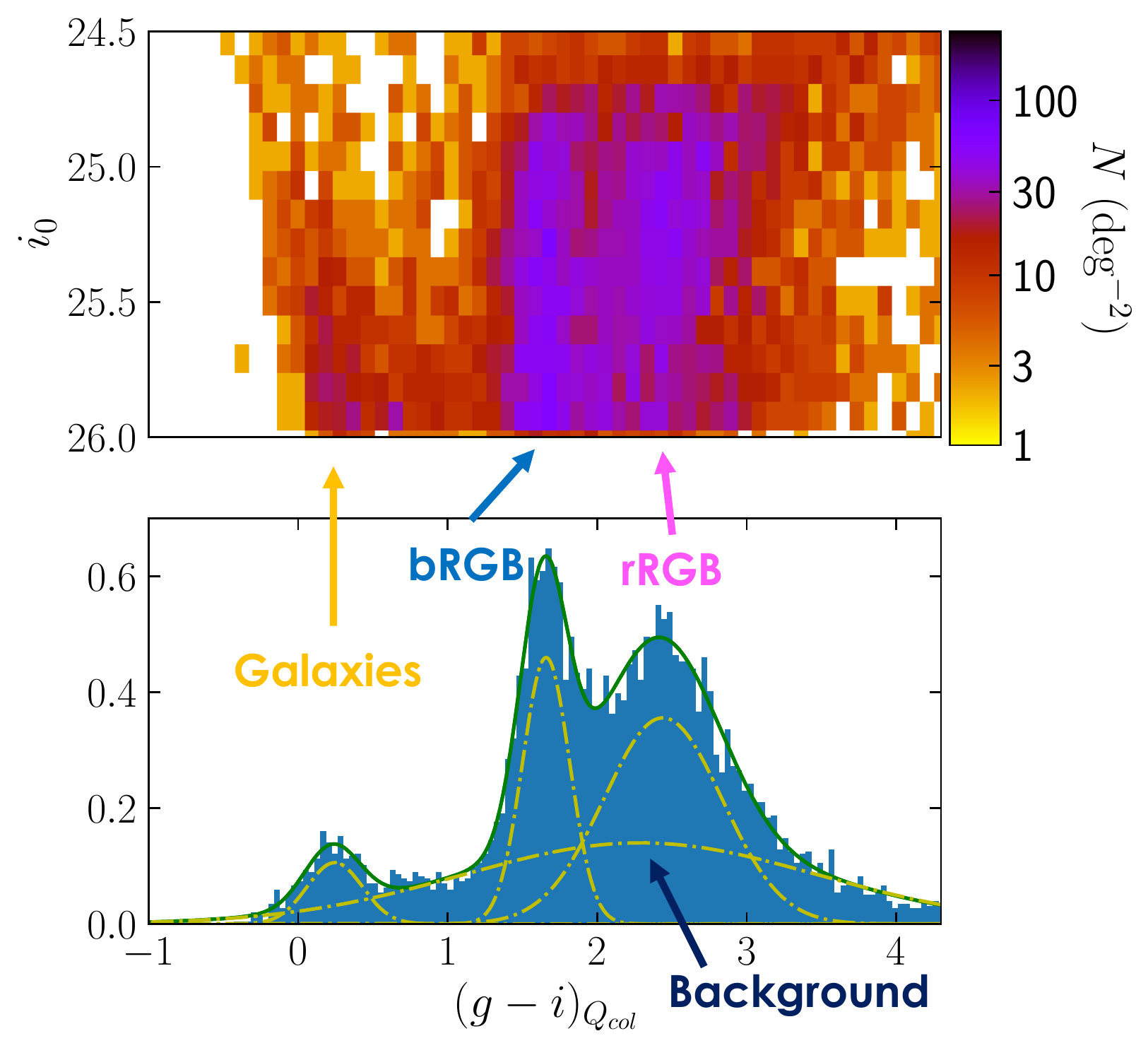}

    \caption{\textit{Top}: A Hess Q-color CMD of stellar sources from 12-30 kpc with magnitudes $24.5\leq i \leq 26$. \textit{Bottom}: A 1D histogram of the $g{-}i$ Q-color (blue) fitted with a four-component Gaussian mixture model. Each Gaussian is shown as a dash--dotted yellow line, with the sum of all four curves displayed as a solid green line. The leftmost peak is the region of unresolved galaxies. The next two peaks represent two different populations of RGB stars, which we call bRGB and rRGB, respectively. The fourth, broad peak is most likely fitting the overextended blue tail or galaxy contamination.}
    \label{fig:gaussian}
\end{figure}

From the distribution of stellar sources between 12-30 kpc from the center of the galaxy, we see that the RGB branch on the CMD looks to have two different regions within it, one centered around a Q-color of $\sim 1.5$ and the other centered around $\sim 2$. We also fit a four-component Gaussian mixture model to the peaks of the 1D Q-color distribution to obtain the peak colors of the two regions. The Q-color CMD and the 1D histogram of the Q-color is shown in Figure \ref{fig:gaussian}. Each fitting Gaussian peak is overlaid in yellow, with the sum of all the peaks shown in green.

A four-component Gaussian mixture model describes the features of the Q-color distribution reasonably well. The RGB region of M94 has two clear and distinct populations corresponding to RGB stars at two different characteristic colors of $Q = (g{-}i)_{TRGB+0.5}$\ = 1.64 and 2.22. In addition, a blue peak accounts for the population of background galaxies; this component has little impact in this study as we select against it in color--magnitude space. The fourth broad peak models (likely imperfectly) the foreground star and compact background population, which has a wide range in color, extending well beyond the colors expected of RGB stars. It is clear from this consideration that fore- and background objects do contaminate our RGB selections, requiring our later use of background subtraction for quantitative investigation of the density and color profiles of RGB stars.

\begin{figure}[!h]
    \includegraphics[width=84 mm]{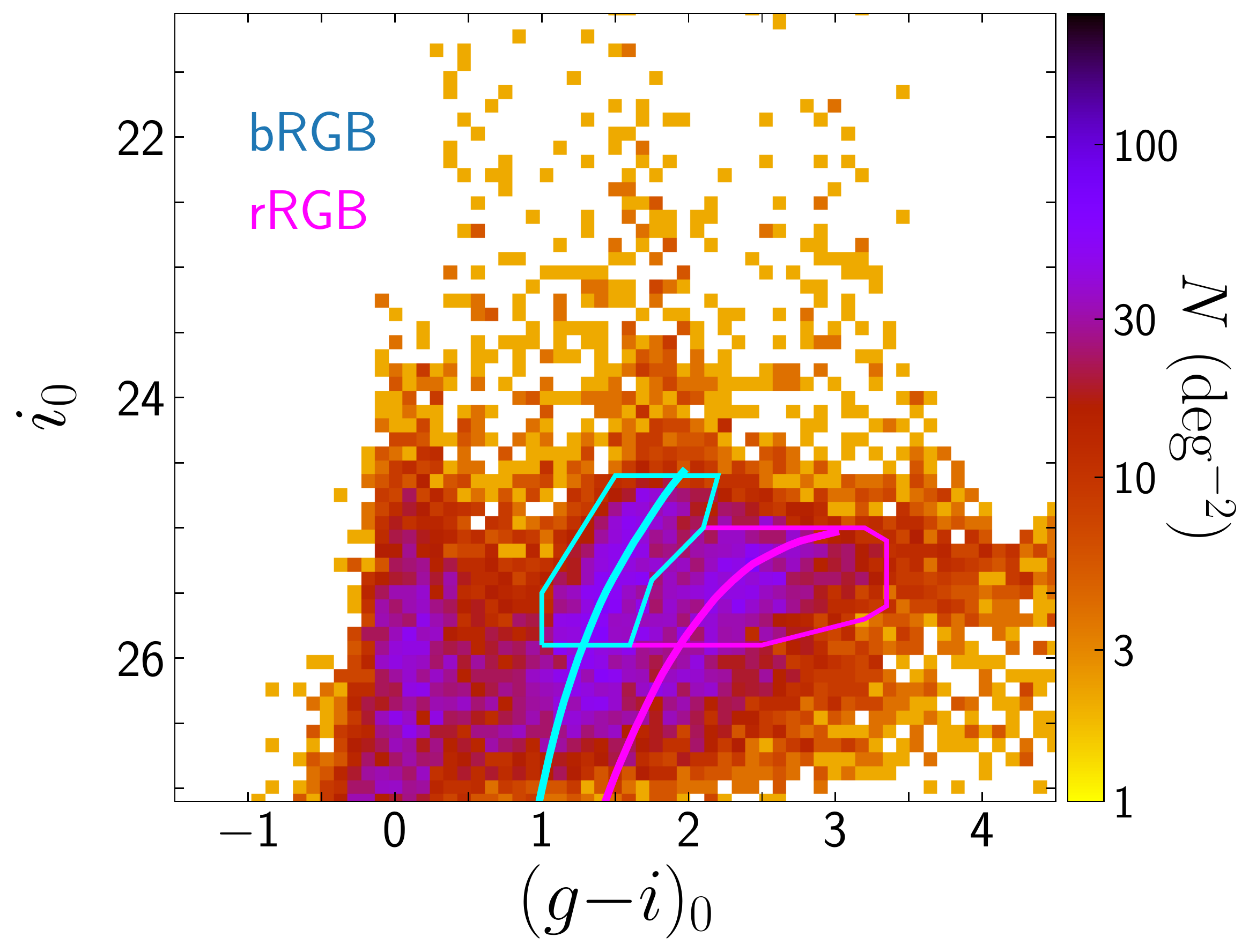}
    \caption{A Hess CMD of stellar sources between 12-30 kpc. Overlaid in cyan and pink are the regions we use for bRGB and rRGB, with their best-fit isochrones from Gaussian fitting and color-metallicity interpolation shown.
}
    \label{fig:hessrgb12}
\end{figure}

Bearing this in mind, we focus on the two RGB stellar populations for the remainder of this paper. These regions are displayed on top of the full stellar source CMD in Figure \ref{fig:hessrgb12}, with best-fit isochrones overlaid. We will refer to the blue-most region as bRGB and the redder one as rRGB throughout. The best-fit isochrones for each region have an  [M/H] = $-$1.4 for bRGB and [M/H] = $-$0.4 for rRGB assuming an age of 10 Gyr. The metallicity for bRGB is derived from generating a set of PARSEC (\citealt{bressan_parsec}) isochrones with metallicities between 0.0001 and 0.02 in 0.00005 increments, finding their TRGB + 0.5 color, and comparing to the peak Q-color of the region that has been transformed to a $g{-}i$ color. The metallicity for rRGB stars is derived star-by-star because the metal-richer isochrones are more tilted and curved than the metal-poorer ones, and so a conversion to Q-color becomes mildly magnitude dependent. For each star, we find the $g{-}i$ color difference at each i-band magnitude for each star in the RGB region. The best-fit metallicity is then taken from the model that minimizes the difference between the $g{-}i$ color of the RGB stars and the isochrone model. Thereby, the median metallicity of the rRGB stars is taken from the median of the individual metallicity measurements of all the rRGB stars\footnote{We note that adopting this process for the bRGB stars gives the same result for bRGB average metallicity as the Q-color based method, [M/H]$_{30 \text{ kpc}}$ = $-$1.4.}.

\moreurgent{We note that it would be possible to fit the RGB colors with different combinations of age and metallicity. In particular, younger, slightly more metal-rich isochrones can also lead to a good fit (we estimate an [M/H] $\sim$ $-$1.07/$-$0.17 and $-$1.25/$-$0.3 for the bRGB/rRGB of a 4 and 7 Gyr population, respectively). \veryurgent{This degeneracy does not affect the main conclusions of this paper, with the degree of systematic error introduced by the choice of stellar population being within the final quoted error bars. Nonetheless, we argue that our choice of a fiducial age for the halo stellar population of 10 Gyr is appropriate. Indeed, magnetohydrodynamics simulations such as those used in \cite{monachesi2019} find that the median age for accreted stellar populations in the stellar halo of Milky Way-mass galaxies is 9.4 Gyr at a galactocentric distance of 30 kpc. Other studies such as by \cite{font2006} and \cite{mccarthy2012MNRAS.420.2245M} also find a lower limit on the age of accreted halo stars to be $\gtrsim10$ Gyr. }}

\begin{figure*}
    \includegraphics[width=0.98\textwidth]{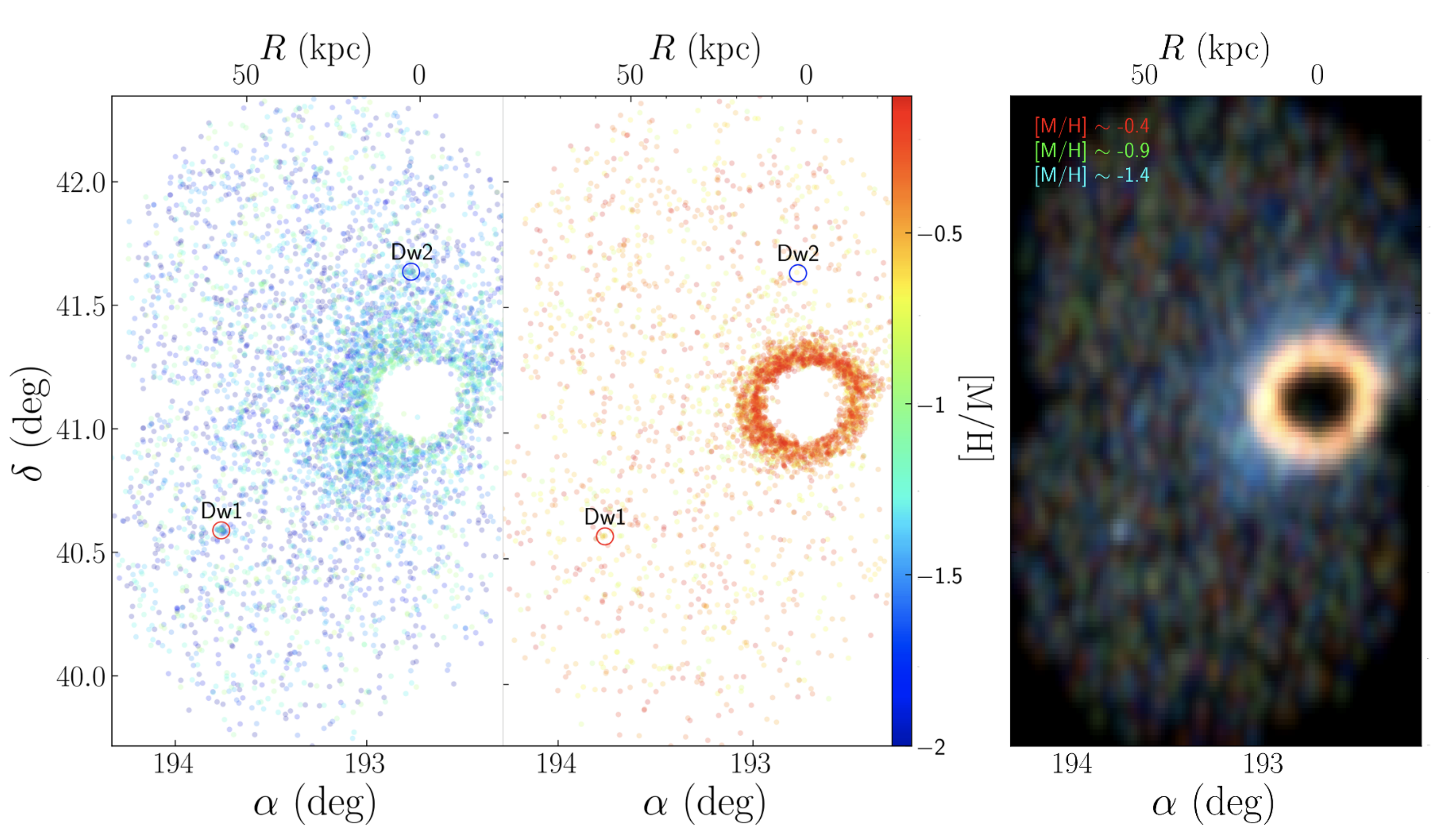}
    
    \caption{Stellar sources in bRGB (left) and rRGB (center), color-coded by their best-fit metallicity. bRGB stars are metal poor and compose M94's outer halo, while rRGB stars are metal rich and comprise most of M94's disk. Also shown are the locations of M94's two known dwarf galaxies, characterized in \cite{Smercina2018}. \textit{Right}: A density image of M94 with each color coded by stars in three metallicity bins: [M/H] $\sim$ $-$0.4 (red), [M/H] $\sim$ $-$0.9 (green), [M/H] $\sim$ $-$1.4 (blue). The map was smoothed using a Gaussian filter and then a square-root scaling was applied. \urgent{All three plots are on the same scale.}}
    \label{fig:rgbboth}
\end{figure*}

In Figure \ref{fig:rgbboth} we show the spatial distribution of stars in bRGB and rRGB color coded by their best-fit metallicities. Superimposed on a roughly uniform background of fore- and background contaminants, we see clear concentrations of stars at M94's distance. This visualization shows the striking difference between the two regions in both spatial distribution and metallicity: bRGB is clearly composed of a more metal-poor population, and its stars are more widely distributed than those in the rRGB population, whose stars are clustered in a dense circular region at small radii until $\sim 30$ kpc. We later calculate the brightness profile of each population, showing that the rRGB population follows an exponential profile that directly follows the outer disk profile from \citet{Watkins2016}; we therefore attribute the rRGB population to M94's disk in what follows. We attribute the bRGB population, which is metal poorer, more diffuse, appears mildly asymmetric, and shows hints of possible streams, to M94's stellar halo.

\begin{figure*}
    \includegraphics[width=0.99\textwidth]{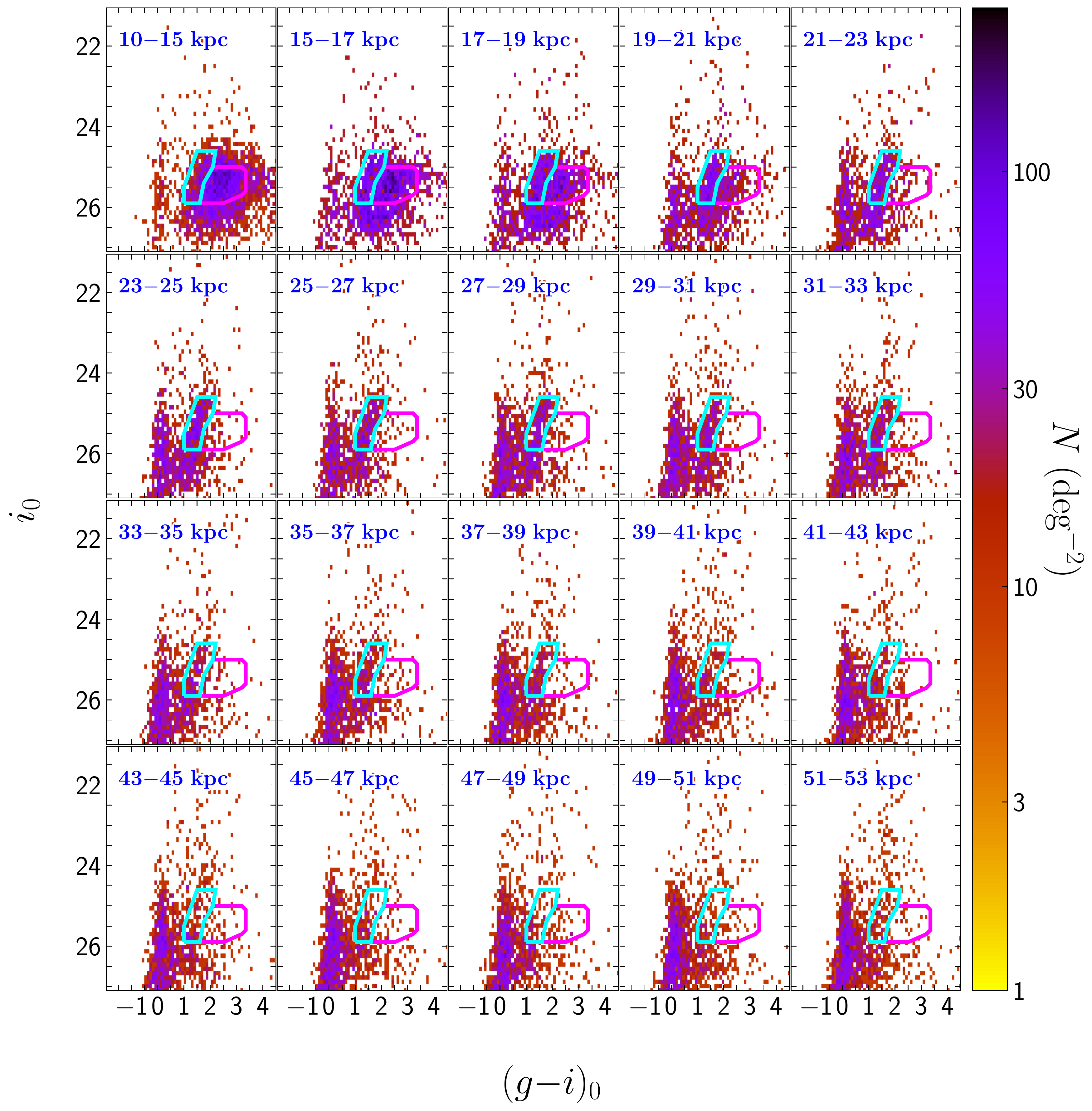}
    \caption{A Hess CMD of stellar sources in various radial bins. bRGB and rRGB selection boxes are shown in cyan and magenta respectively.}
    \label{fig:radial_hess}
\end{figure*}

In order to calculate the color and surface brightness profiles accurately, it is important to determine the radius at which we are dominated by the roughly uniform background of contaminants apparent in Figs.\ \ref{fig:gaussian} and \ref{fig:rgbboth}. We examine Hess CMDs covering a series of radial bins from the center of the galaxy, as shown in Figure \ref{fig:radial_hess}.

From the visual inspection, we conclude that the bRGB region starts to be dominated by background galaxies after $\sim 50$ kpc, and the rRGB region after $\sim 30$ kpc. After these thresholds, the density counts drop drastically in each respective region and we cannot see a clear RGB feature. These background thresholds are used for the rest of the analysis.

\subsection{Artificial Star Tests} \label{sec:ast}

Before we can measure these profiles however, we need to account for completeness and crowding in our sample. As illustrated in Figure \ref{fig:rgbboth}, there is a gap in the very central part of the galaxy due to crowding effects such that the HSC reduction pipeline was unable to extract sources reliably. There are also sources that overlap in high-density regions, especially around the first few radial bins away from the galaxy's center. The HSC source detection pipeline is not able to locate and categorize every source with full accuracy, especially those that are blended together, making it necessary for us to account for this source of error. 

In order to quantify the completeness of our sample rigorously, we ran a suite of artificial star tests (ASTs). In order to avoid the injected artificial stars from causing crowding, we inject a total of 674,980 artificial stars split into five runs of 134,996 stars each. Artificial stars were drawn from uniform grids in position, in $i$-band magnitude between $i=22.5$ and $i=28$, and color between $g{-}i = -2$ and $g-i = 5$ colors. The $r$-band magnitudes were then calculated assuming each artificial star lies on the $g{-}r$/$r{-}i$\ stellar locus. At the input artificial source density of $\sim 10$\,arcmin$^{-2}$, these artificial sources do not measurably effect the measurement of the source properties by the pipeline \citep[e.g.,][]{Smercina2020}. The pipeline was then run on each of these five artificial star runs in the exact setup used to process the original observations, resulting in a list of recovered artificial stars. For our purposes, an input artificial star was declared recovered if its recovered $i$-band magnitude was within 0.5\,mag and its recovered position within $0\farcs3$ of its input magnitude and position, respectively.

Our particular interest is in calculating the completeness in each of our two RGB regions (bRGB and rRGB), in the same regions of the image used to create the star count profiles (e.g. Fig.\ \ref{fig:SBmassprofile}). We choose a set of radial bins, 5 kpc wide from 10-15 kpc and 2 kpc wide thereafter. Artificial stars in the color--magnitude regions bRGB and rRGB were analysed for each spatial region, and the fraction of those recovered was weighted by the ratio of the number of input stars to the expected luminosity function of RGB stars with their appropriate best-fit isochrones (from Fig.\ \ref{fig:hessrgb12}), as a function of $i$-band magnitude. In this way, for each spatial region of interest, a single completeness for each of bRGB and rRGB stars was determined with which to scale the observed number of bRGB and rRGB stars, respectively, in each region.

\section{Results}
\label{sec:results}
Here we present the $g{-}i$ color and median surface brightness (SB) profiles of M94's RGB. We then use this to derive a stellar mass of M94's halo and infer the mass accreted during its most massive-major merger.

\subsection{Surface Brightness Profile}
\label{sec:sbprofile}

For studying resolved stellar populations in the context of mergers, typically one would like to measure only the stars along the minor axis of a galaxy since these are relatively free of contamination from in-situ stars (see \citealt{Monachesi2016b, Smercina2020}). Since M94 is not as highly inclined as galaxies such as M81, its minor axis traces stars in the plane of the disk, so we once again used the same circular radial regions as referenced in Section \ref{sec:ast} for our analysis. We choose to analyze radial profiles out to a 70\,kpc radius. For the bRGB (halo) population we adopt a background region from a 50-70\,kpc projected radius, yielding a background density of 0.21 arcmin$^{-2}$. The rRGB (disk) is detectable only at a much smaller radius, and a background density of 0.13 arcmin$^{-2}$, determined between 30-70\,kpc, was adopted. 

For each radial bin, we calculate the RGB-selected and background-subtracted source density, dividing by the completeness fraction found from our AST analysis. The conversion between the number of RGB stars in the observed magnitude range to stellar mass assumes a [M/H] = $-1.4$ ($-0.4$) isochrone for the bRGB (rRGB) respectively and a 44\% mass loss between initial and present day mass which is compatible with a Chabrier IMF. When calculating surface brightnesses, we adopt a $V$-band mass to light ratio of 1.25, appropriate to $\sim 10$\,Gyr-old populations with the range of relevant metallicities from \citet{bruzalcharlot2003} single-burst models. 

\begin{figure*}
    \centering
    \includegraphics[width=0.99\textwidth]{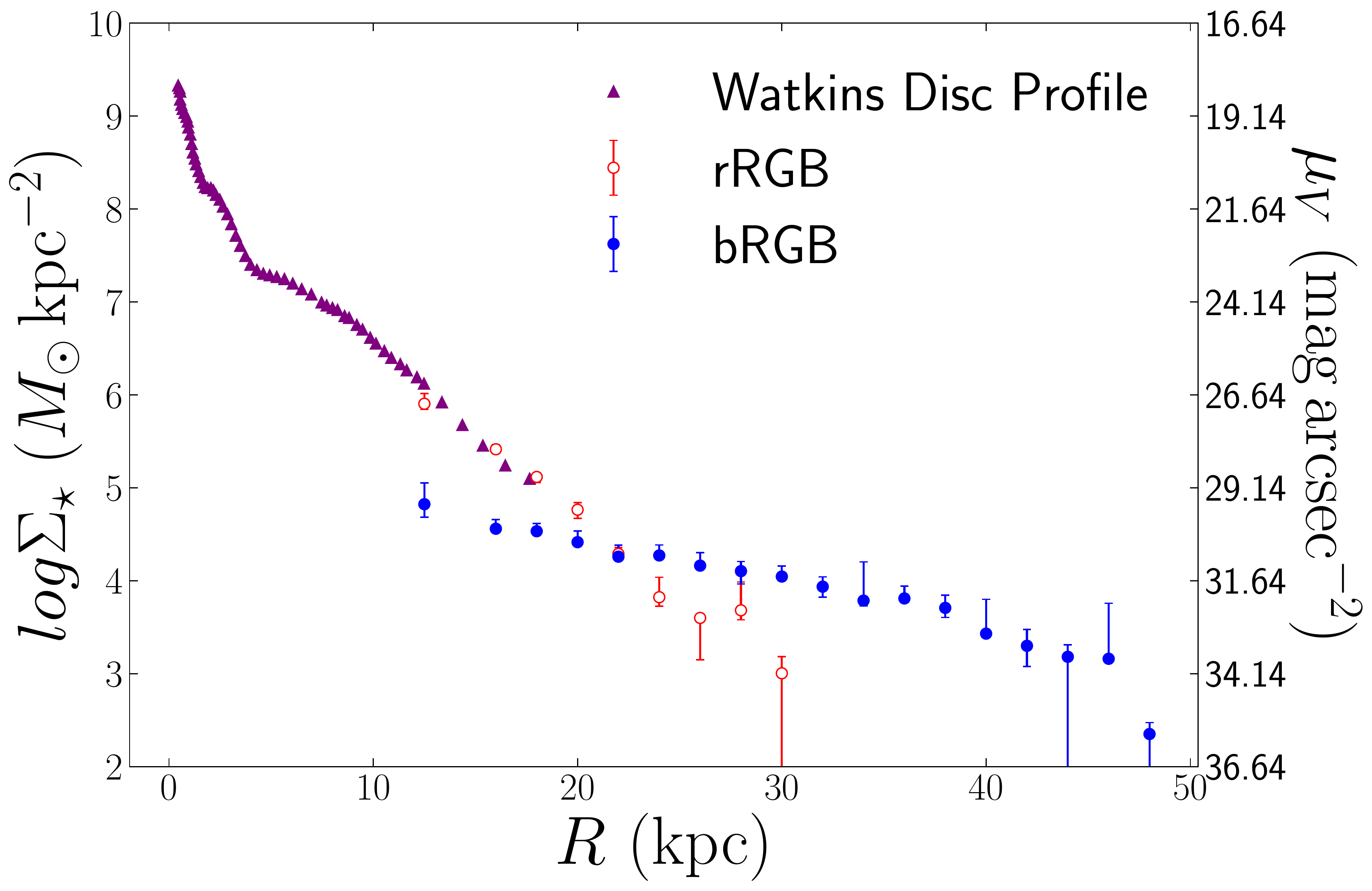}

    \caption{The surface mass density and surface brightness profile of stars in different radial bins for bRGB (blue closed circles) and rRGB (red outlined circles). Plotted in purple is the surface brightness profile of M94's outer disk as calculated by \cite{Watkins2016}, translated to V-band magnitude.}
    \label{fig:SBmassprofile}
\end{figure*}

Formal uncertainties from counting statistics are negligible.  Instead, we estimate uncertainties in our density profiles that incorporate larger-scale variations (owing to halo substructure or variations in background) by assessing the variation of our profiles as a function of azimuth. We divide the dataset into radial slices of 45 degrees wide and repeat the same background estimate and mass density calculations as before for both RGB regions. At each radius bin, we perform a bootstrap resampling analysis and chose one of the quadrant mass densities at that radius, with replacement. We repeat this 1000 times and find the 16th and 84th percentile values of the resulting distribution, which we use as the upper and lower error bars on Fig.\ \ref{fig:SBmassprofile}.

\begin{figure}
    \centering

    \includegraphics[width=0.99\columnwidth]{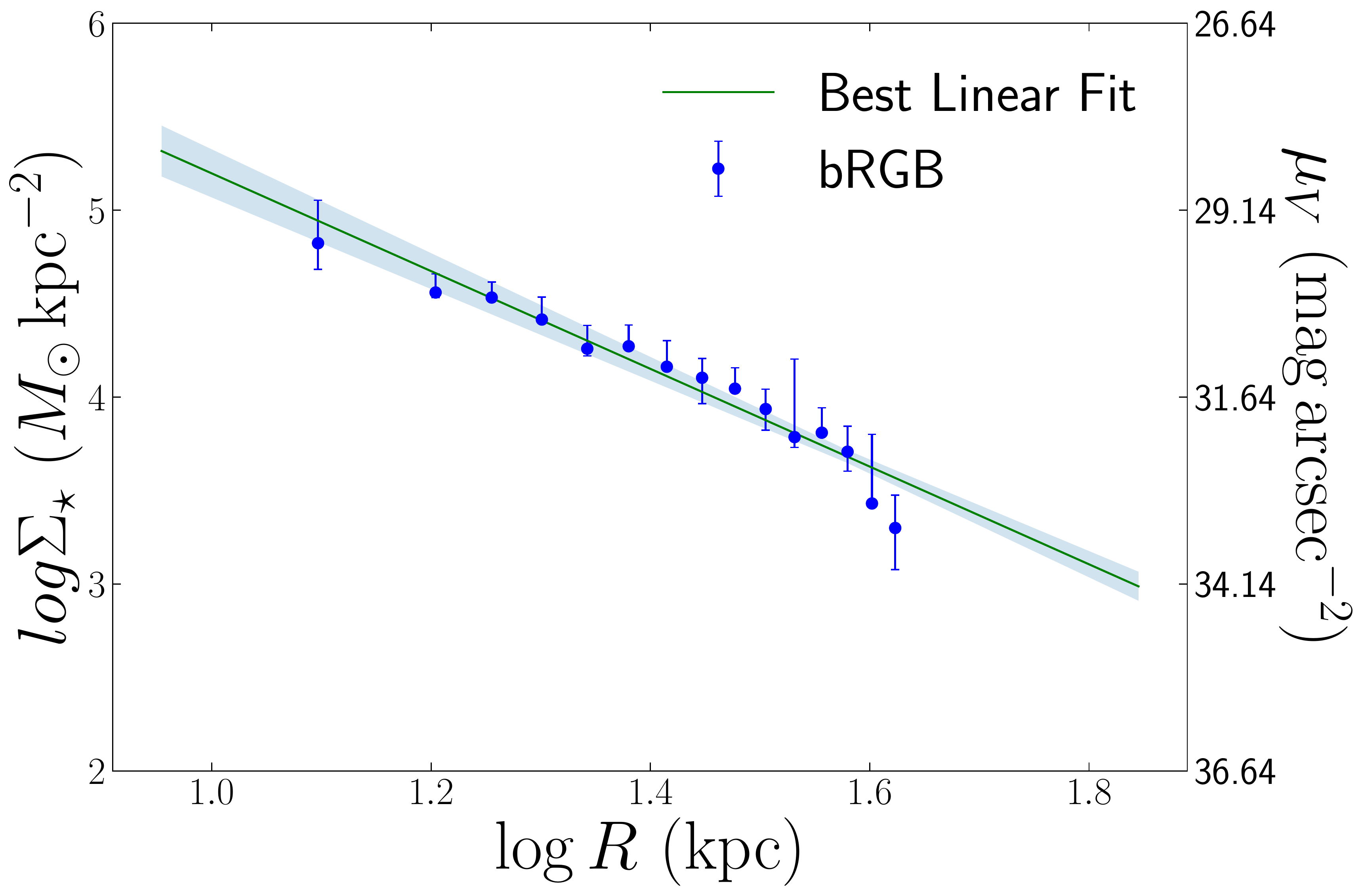}

    \caption{Best linear fit to the log-log mass density profile for data between 16-42 kpc.}
    \label{fig:linearfit}
\end{figure}

We also overplot the mass density from \cite{Watkins2016}, who measured M94's inner disk. These papers assume elliptical annuli around M94's center, while we used circular annuli, so we adjust their given radii for this and convert to equivalent circular radii, assuming an axis ratio of b/a = 0.79 \citep{Watkins2016}. This is shown in Figure \ref{fig:SBmassprofile}. The rRGB profile agrees closely with the profile of the outer disk from \citet{Watkins2016} within 20\,kpc, the range where the measurements overlap. We can also see that bRGB's profile is much flatter than rRGB's. These stark differences leads us to affirm that bRGB and rRGB represent two different stellar populations in the galaxy. rRGB shows an exponential drop in its profile and agrees with \cite{Watkins2016}'s inner disk profile, strongly suggesting that its stars are actually in the disk of M94, while bRGB represents the halo population.

To measure the total accreted \urgent{stellar mass}, we fit a three-parameter power law model with intrinsic scatter following \citealt{hogg2010} from 16 -- 42 kpc (Fig. \ref{fig:linearfit}). We use a power law of the form: $\Sigma = a \log_{10}(\frac{x}{R_0})+ \Sigma_0$ with a likelihood of $$log \mathcal{L} = -0.5 \sum_{i=0}^n \frac{(\Sigma_i-\Sigma(x,\Sigma_0))^2}{\sigma^2} + log({\sigma^2}),$$ where: $$\sigma^2 = \sigma_\Sigma^2 + V,$$ with $\sigma_\Sigma$ being the error in the mass density in each radial bin and $V$ being the variance of the intrinsic scatter. $R_0 = 37.9$ kpc is a normalization factor that we take to be the mean of the radial bins used for fitting. We find best fit parameters of $a= -2.61^{+0.16}_{-0.17}$, $\log \Sigma_0= 3.68^{+0.03}_{-0.03} M_\odot {\rm kpc}^{-2}$, Following \cite{Harmsen2017}, we integrate this profile from 10-40 kpc and calculate an accreted stellar mass of $M_{*, 10-40} = 9.25 \times 10^7 M_\odot$. \moreurgent{\cite{Harmsen2017} uses simulated stellar halo data from \cite{bullockjohnson} to estimate that the total accreted halo mass, $M_{acc}$, is about 3 times that of the mass between 10-40 kpc, $M_{*, 10-40}$, with variation of $\sim$40\% from halo to halo. Accounting for this uncertainty in our error calculation, we extrapolate to total accreted mass following \moreurgent{this method} and get a total accreted \urgent{stellar} mass of $M_* = 2.77^{+1.54}_{-1.00}\times10^8 M_\odot$.}

\subsection{Color Profile of the bRGB halo population}
\begin{figure}
    \centering
    \includegraphics[width=0.99\columnwidth]{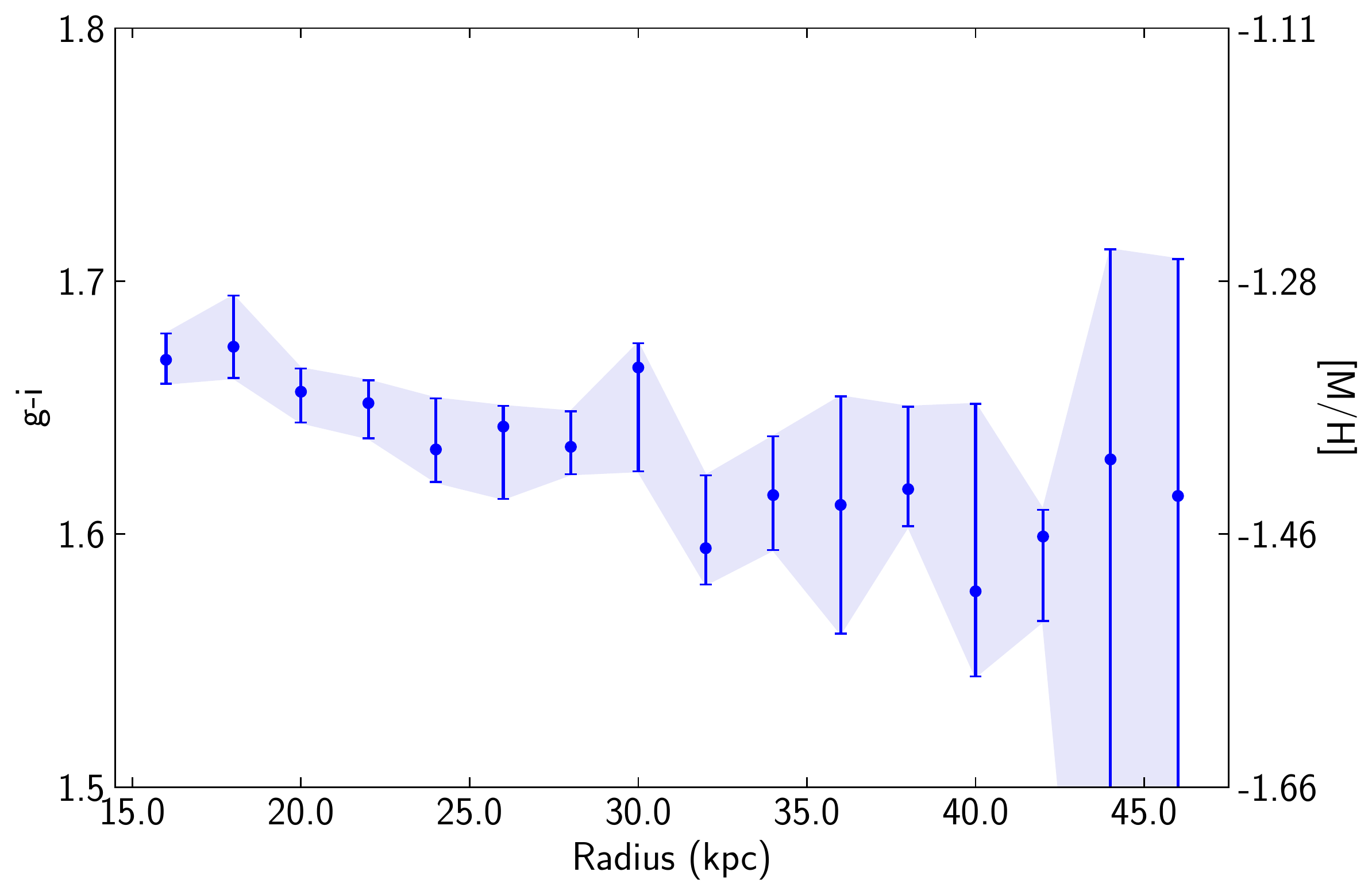}

    \caption{The $g{-}i$ color profile of resolved stars in the bRGB region. Metallicity, which is calculated from the color and set of isochrones previously mentioned in \ref{sec:rgbregions}, is shown along the right y-axis. We estimate a relatively mild halo metallicity profile over the halo from 16-40 kpc of $-0.005$ dex/kpc.}
    \label{fig:color}
\end{figure}

We calculate a median $g{-}i$ color and metallicity profile for bRGB using the same bins as the mass density profile. We first calculate a 2D histogram of all the sources and then estimate the area and sources in the background region (everything between 50 - 70 kpc) to get the background source color distribution. For each radial bin, we find the cumulative distribution of background Q-color weighted by the ratio of the bin area to background area, which we use to get the total number of background sources that correspond to a certain color. We then find the cumulative distribution of bRGB stars in each bin and subtract the cumulative background distribution normalized to the area of the radial bin and find the median of this background-subtracted distribution, converting back to $g{-}i$ color. The resulting color profile is shown in Figure \ref{fig:color}. We estimate a metallicity gradient of slope $\sim-0.005$ dex/kpc between 16 and 40 kpc.

The color gradient is $d(g-i)/dr = -0.0028$  mag/kpc. 
Assuming 10\,Gyr old stellar populations, and recalling that RGB color is a very weak function of age, the $g-i$ color should correspond to the metallicity of the population, shown schematically on the right-hand side of Fig.\ \ref{fig:color}. While the mapping to metallicity is non-linear, in the color range relevant to M94's bRGB population a metallicity gradient of $d{\rm [M/H]}/dr \sim -0.005$ dex/kpc is implied, if the $g-i$ color gradient is interpreted as being due to metallicity alone. To compare this with other halos requires translating to the HST passbands used in \citet{Monachesi2016} and \citet{Harmsen2017}. A linear fit to $g-i$ vs. $F606W-F814W$ for stars with HST measurements in the rough color range around $g-i \sim 1$ gives a fit $g-i \sim 1.91 (F606W-F814W) - 0.64$; correspondingly, the implied color gradient in $F606W-F814W$ is $d(F606W-F814W)/dr = -0.0015$ mag/kpc. This color gradient is rather mild, similar to galaxies like the Milky Way, NGC 253 or NGC 4945.

\section{Discussion}

\subsection{M94's Stellar Halo}
\begin{figure*}
    \centering
    \includegraphics[width=0.75\textwidth]{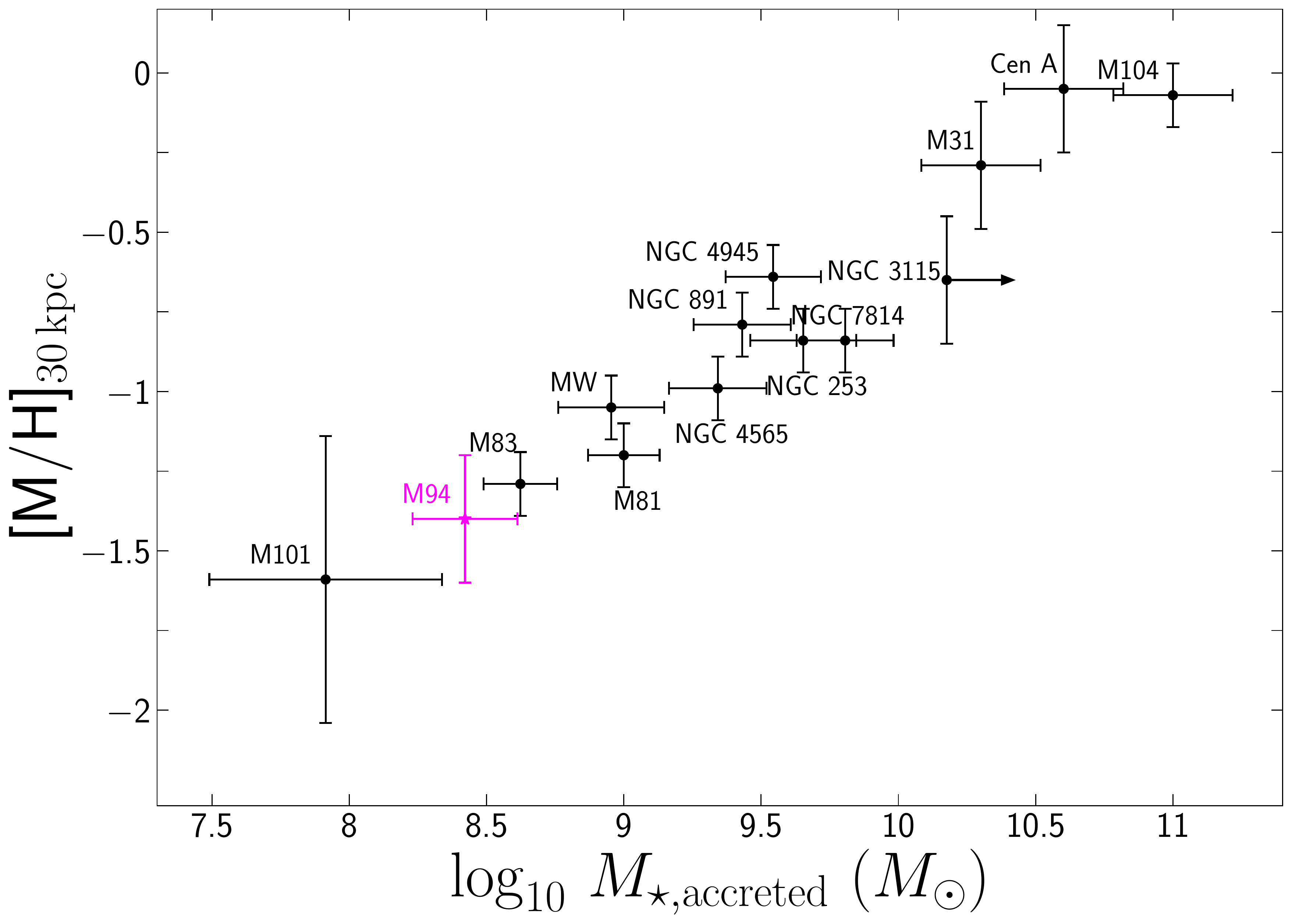}

    \caption{[M/H] at 30 kpc from the center as a function of the total stellar mass accreted for various galaxies in the Local Universe, reproduced from \citealt{Smercina2021}. Galaxies whose metallicities were listed as [Fe/H] were converted to their respective [M/H] following \cite{streich}. M94 falls along the general trend, having one of the smallest and most metal-poor halos of any galaxy in the Local Universe.}
    \label{fig:massmetallicity}
\end{figure*}

We follow numerous past works and use M94's stellar halo properties to make inferences about its merger history, e.g. \cite[][and references therein]{Bell2017}. We show M94's halo in context of the observed stellar halo metallicity--mass relation for 14 nearby MW-\urgent{stellar} mass galaxies in Figure \ref{fig:massmetallicity}, which was first discussed in \cite{Harmsen2017}. Since RGB color is affected primarily by [M/H] and not [Fe/H] \citep{streich}, we use the $\alpha$-agnostic [M/H] instead of converting our measurement to [Fe/H]. M94 follows the halo metallicity--mass correlation well and contains one of the least massive and most metal-poor halos compared to these other disk galaxies, with only M101 harboring a smaller, more anemic halo. This generally agrees with the general expectations from theoretical accretion-only models wherein the halo is accreted from the merger of its last satellite progenitors, and that a smaller halo is formed through the aggregation of a few less-massive satellites instead of being dominated by a single very massive merger \citep[][]{monachesi2019}. This also points to M94 likely having a more quiescent and quiet merger history compared to the majority of local MW-\urgent{stellar} mass galaxies \citep{deanson2015}.

\cite{DSouzaBell2018a} found a correlation between the mass of a galaxy's dominant merger, $M_{dom}$, and the halo's total accreted \urgent{stellar} mass, $M_{acc}$, (see also \citealt{zhu2022mass}). \moreurgent{This assumed that only a fraction, \text{frac}$_\text{Dom}$, of the accreted mass is contributed to the stellar halo by the dominant merging partner. For the majority of galaxies that fall a slight ways below a 1-to-1 line on the $M_{dom} - M_{acc}$ plane, \text{frac}$_\text{Dom}$ is less than 1. In most cases, the accreted halo mass $M_{acc}$ is greater than the mass of the dominant progenitor $M_{dom}$. This is especially true at the lower mass end of $M_{acc}$ and indicates that the halo contains contributions from a number of other merger events, consistent with our idea of the hierarchical build-up of galaxies.} In order to estimate \moreurgent{the mass of M94's dominant merger partner}, we follow \citet{DSouzaBell2018a} and correlate the dominant merger mass with the total accreted \urgent{stellar} mass for galaxies taken from the Illustris simulation suite \citep{Genel2014,Vogel2014_Nature,Vogel2014_MNRAS}, chosen to have dark matter halo masses of $11.7 \leq \log M_{DM} \leq 12.15$.  We fit this distribution with a line, and infer the mass of M94's dominant merger from our accreted mass calculated in Section \ref{sec:sbprofile}. The uncertainty in the dominant mass estimate is taken as a sum in quadrature of the intrinsic scatter in the distribution and the uncertainty of the measured accreted mass. The resulting dominant \urgent{stellar} mass is \moreurgent{estimated to be} $\log M_{dom} = 8.15^{+0.29}_{-0.22} M_\odot$. The Small Magellanic Cloud's (SMC) stellar mass is $4.6\times10^8 M_\odot$ (\citealt{smcmass}); M94's most massive merger partner is likely to be less likely than even the SMC.

\begin{figure}
    \centering
    \includegraphics[width=0.89\columnwidth]{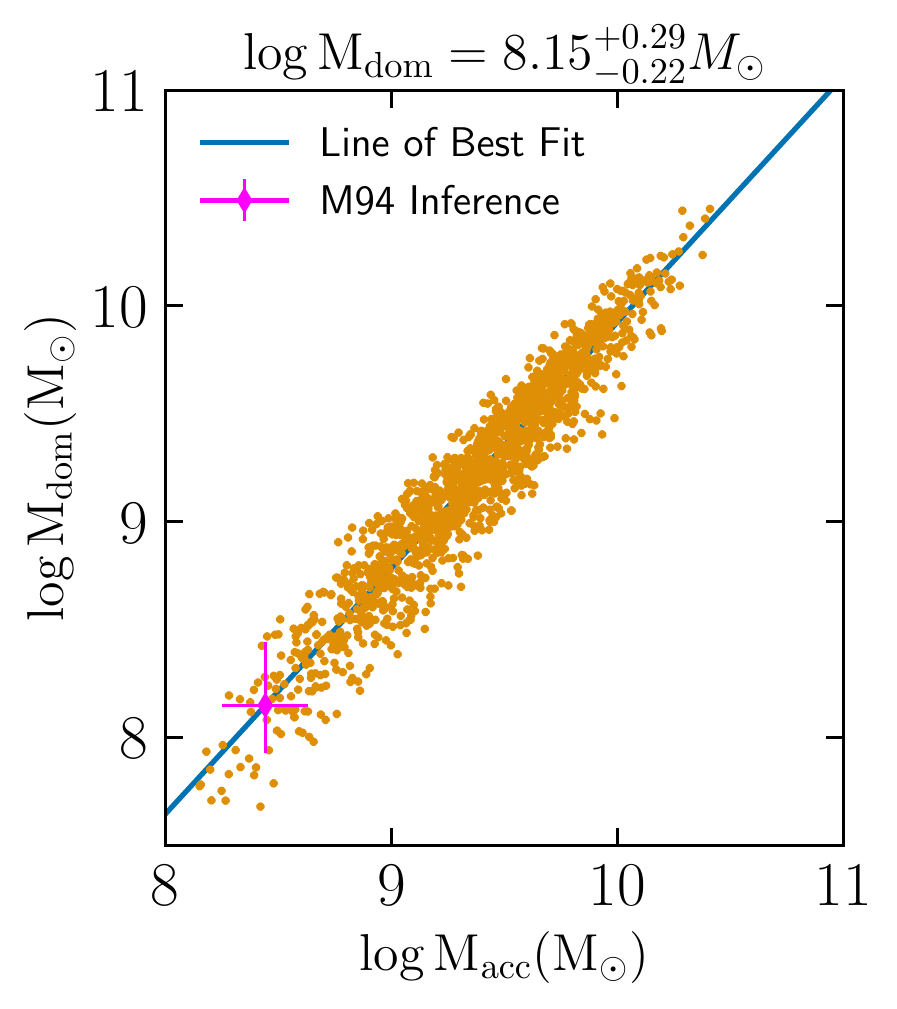}
    \caption{
    Orange points show the masses of the dominant progenitor galaxies as a function of the total accreted stellar mass for galaxies in dark matter halos with masses between $11.7 \leq \log M_{DM} \leq 12.15$ from the Illustris simulation. The line of best fit to the data is shown in blue and M94's inferred location is displayed as a pink diamond.}
    \label{fig:maccmdom}
\end{figure}

\subsection{Psuedobulge Inferences}

Thus we see that M94 is characterized by an inactive merger history which gives rise to its small, metal-poor halo. Two other nearby galaxies, M83 ($M_{halo} = 4.2 \times 10^8 M_\odot$, Cosby et al., in prep) and M101 ($M_{halo} =8.2\times 10^{7} M_\odot$, \citealt{Jang2020}), exhibit this same type of quiet merger history. Yet all three of these galaxies have very different bulge masses.  While M94's pseudobulge is massive (\citealt{Trujillo_2009}, almost $10^{10.5} M_\odot$), 
there is no evidence for a classical bulge in either M101 \citep{KormendyDrory_bulgelessgiants_2010} or M83 (\citealt{dejong2008}, \citealt{fisherdrory}); both galaxies hosting small ($\le 10^{9.5} M_\odot$) pseudobulges. These three galaxies --- with outwardly very similar merger histories --- have such wildly dissimilar bulge properties, suggesting that the mass of a galaxy's dominant merger is not simply correlated with the mass or properties of its bulge.

\begin{figure}

    \centering
    \includegraphics[width=0.99\columnwidth]{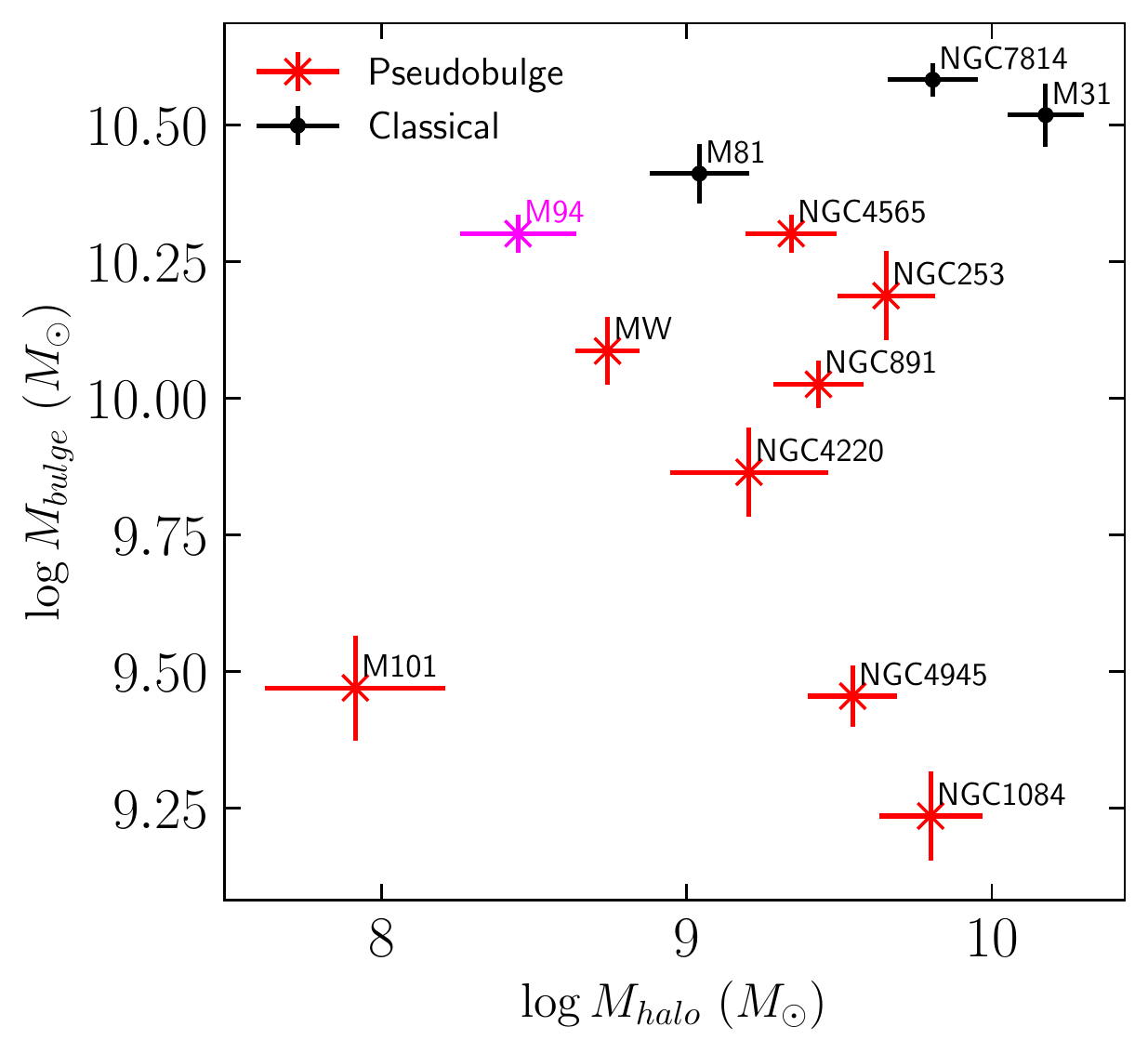}

    \caption{The accreted halo mass as a function of bulge mass for galaxies in the Local Universe where both measurements exist. M94 is highlighted in pink text.}
    \label{fig:mbulgemhalo}
\end{figure}

We connect M94 to the broader population of galaxies with measured halos and bulges in Figure \ref{fig:mbulgemhalo}, with galaxies marked as having either a classical bulge or a pseudobulge. If we take the halo \urgent{stellar} mass to act as a proxy for the mass of a galaxy's dominant merger, then we see that almost every other pseudobulge-hosting galaxy in our Local Universe underwent a larger merger than M94, and yet hosts less massive bulges. This adds evidence to the picture that mergers are not the primary drivers of pseudobulge creation, and any role that they play in determining its properties would be complicated and intertwined with other, secular processes that are on-going within the galaxy \citep[][]{Bell2017, gargulio2019, gargulio2022}. \urgent{Indeed, gas flows in so-called ``oval distortions" like the kind M94 has can be a major impetus for the formation of pseudobulges \citep{speights2019_ovaldistortion} and can also fuel star-forming rings and enhance star-formation in general \citep{athanaousla1984BAICz..35..380A, mulderandcombes1996, Li2016}. M94 currently has about equal fractions of intermediate-age and old-age stars in its pseudobulge, found by taking spectra within its central 1'' x 1.1'' region \citep{cidfernandes2004, mason2015}, indicating semi-recent star-formation, potentially from gas driven inward from a Lindblad resonance \citep{wong2001, Trujillo_2009}.}

\moreurgent{What makes M94 even more interesting is features such as its warped disk, with its HI disk misaligned with its optical counterpart and differential alignment of its inner disk (oval distortion) and outer disk, and its highly non-circular motions \citep{deBlok2008, walter2008_thingsnoncirc, Trujillo_2009}. \cite{Trujillo_2009} rules out merger as an explanation for M94's differential inclination, citing merger simulations that give too low of a resulting inclination to match observations. Instead, \cite{Trujillo_2009} note that the inclination discrepancy in this case goes away if the ``inner disk" is an oval distortion, so it has a completely different ellipticity than the outer disk. And while merger is a common explanation for irregularities in galactic morphology, given our inferences regarding M94's quiet merger history, the presence of these features indicate that the gas morphology of a galaxy can be strongly distorted by processes that involve no merger with an established galaxy. M94 would not be the only curious example -- galaxies such as M83 have been found to also harbor a warped and disturbed disk \citep{Barnes2014, Heald2016}, yet have no sign of a recent large merger event (Cosby et al., in prep).}

\subsection{Satellite Populations}

The measure of M94's accreted and dominant merger masses has also been used to gain insight into the satellite populations of MW-\urgent{stellar} mass galaxies. Satellite luminosity functions have now been measured for a substantial number of Local Volume galaxies, finding a large range in the number of satellites in each group (\citealt{Smercina2018}, \citealt{Muller2019}, \citealt{Carlsten2020}; see \citealt{Carlsten2022} for an overview), where M94 is a particularly extreme example, having only two classical satellites \citep{Smercina2018}.  The number of satellites appears to vary with the host galaxy's stellar mass \citep{Carlsten2020} as would be expected with a scaling with overall dark matter halo mass. Yet, considerable scatter remains, and has been claimed to correlate with several different measures that relate to a group's merger history: `tidal index' --- a measure of enviromental density (D.\ Sand, priv. comm.), magnitude gap \citep{Kim2022}, and the mass of the galaxy's most massive past merger \citep{Smercina2021}. While simulations do not yet show such behavior \citep{Smercina2021}, such correlations may broadly reflect the delivery of satellites in large merger events \citep{Deason2015,dsouzabell2021}. Owing to M94's low number of satellites, it has particular constraining power in such correlations, and M94's combination of low halo mass (thus low-mass dominant merger) and low satellite number are consistent with the view that satellite number correlates with merger mass \citep{Smercina2021}.

\section{Conclusions} 

In this paper, we characterize the stellar halo of the spiral galaxy M94 (NGC 4736). M94 is unique in that it contains the largest pseudobulge in the Local Universe. The evolution of central structures such as bulges and pseudobulges hinge on the interplay of both secular and non-secular processes. Therefore, we use M94's stellar halo as a fossil record of its previous merger history to infer how much impact it could have had on the formation of M94's massive pseudobulge.

We use deep, high resolution data from Subaru's Hyper Suprime-Cam (HSC) to distinguish between stars and galaxies using both size and color--color criteria. The color--magnitude diagram of these sources showed two clear red giant branches --- one relatively blue (bRGB) and one substantially redder (rRGB). After correcting for completeness, the surface density profile of the rRGB is consistent with that of M94's outer disk \citep{Watkins2016}, while the bRGB has a profile and structure consistent with a metal-poor stellar halo.

Integrating the halo profile from 10 -- 40 kpc, we estimate a total accreted \urgent{stellar} mass of $M_* = 2.8^{+1.5}_{-1.0}\times10^8 M_\odot$. Using the relation between $M_{dom}$ and the accreted mass from \citet{DSouzaBell2018a}, we \moreurgent{infer} that the mass of M94's dominant merger is roughly $\log M_{dom} = 8.15^{+0.29}_{-0.22} M_\odot$, which is smaller than the SMC, indicating that M94 \moreurgent{likely} underwent a very quiet merger history compared to other local galaxies such as the MW, M81 or M31. We also calculated a color and metallicity profile of the halo, showing that M94 hosts one of the most metal-poor MW-\urgent{stellar} mass galaxy halos in the Local Universe (comparable to only M101), with a median metallicity at 30 kpc to be [M/H] $\simeq -1.4$. Therefore, we infer that M94’s pseudobulge \veryurgent{and other interesting morphological features }were not \veryurgent{significantly impacted} by its most major merger. In fact, every other galaxy in the Local Universe that has a pseudobulge hosts a smaller pseudobulge than M94’s, despite having undergone a larger merger, indicating that other processes are most likely the major stimuli for its creation and growth. \veryurgent{Given our inferences, this would also have }larger implications on galaxy formation and evolution: we now \veryurgent{theorize} that M94, M101, and M83 underwent \veryurgent{similarly} inactive merger histories\veryurgent{, with very low-mass dominant mergers,} and yet the three galaxies have completely different central structures, \veryurgent{in particular their pseudobulges. This supports} the picture that that there is a more complex mechanism which determines a galaxy’s structural properties and that merger history cannot always reliably indicate what structure a galaxy will form.

\section{Acknowledgements}

This work was partly supported by the National Science Foundation through grant NSF-AST 2007065 and by the WFIRST Infrared Nearby Galaxies Survey (WINGS) collaboration through NASA grant NNG16PJ28C through subcontract from the University of Washington. AM gratefully acknowledges support by FONDECYT Regular grant 1212046 and by the ANID BASAL project FB210003, as well as funding from the Max Planck Society through a "PartnerGroup” grant. This research has made use of NASA's Astrophysics Data System Bibliographic Services. 

Based on observations utilizing Pan-STARRS1
Survey. The Pan-STARRS1 Surveys (PS1) and the PS1 public science archive have been made possible through contributions by the Institute for Astronomy, the University of Hawaii, the Pan-STARRS Project Office, the Max-Planck Society and its participating institutes, the Max Planck Institute for Astronomy, Heidelberg and the Max Planck Institute for Extraterrestrial Physics, Garching, The Johns Hopkins University, Durham University, the University of Edinburgh, the Queen’s University Belfast, the Harvard-Smithsonian Center for Astrophysics, the Las Cumbres Observatory Global Telescope Network Incorporated, the National Central University of Taiwan, the Space Telescope Science Institute, the National Aeronautics and Space Administration under Grant No. NNX08AR22G issued through the Planetary Science Division of the NASA Science Mission Directorate, the National Science Foundation Grant No. AST-1238877, the University of Maryland, Eotvos Lorand University (ELTE), the Los Alamos National Laboratory, and the Gordon and Betty Moore Foundation. 

Based on observations obtained at the Subaru Observatory, which is operated by the National Astronomical Observatory of Japan, via the Gemini/Subaru Time Exchange Program. We thank the Subaru support staff for invaluable help preparing and carrying out the observing run. 

The authors wish to recognize and acknowledges the very significant cultural role and reverence that the summit of Maunakea has always had within the indigenous Hawaiian community. We are most fortunate to have the opportunity to conduct observations from this mountain.

\facilities{Subaru Observatory/HSC} 

\software{\code{HSC Pipeline} (\citealt{HSC_pipeline}), \code{Python 3.9} (\citealt{python}) \code{— Matplotlib} (\citealt{matplotlib}), \code{Numpy} (\citealt{numpy}), \code{Scipy} (\citealt{scipy}), \code{Astropy} (\citealt{astropy}), \code{Jupyter} Notebooks (\citealt{jupyter})}

\bibliographystyle{aasjournal}
\bibliography{sample63}



\end{document}